\documentclass[a4paper]{article}

\usepackage{INTERSPEECH2022}
\usepackage{subfigure}
\usepackage{multirow}
\usepackage{array}
\usepackage{color}
\newcolumntype{P}[1]{>{\centering\arraybackslash}p{#1}}
\usepackage{url}
\usepackage{hyperref}

\title{Joint Noise Reduction and Listening Enhancement \\ for Full-End Speech Enhancement}
\name{Haoyu Li$^1{^,}^2$, Yun Liu$^1{^,}^2$, Junichi Yamagishi$^1{^,}^2$}

\address{
  $^1$National Institute of Informatics, Japan\\
  $^2$The Graduate University for Advanced Studies (SOKENDAI), Japan}
\email{haoyuli@nii.ac.jp, yunliu@nii.ac.jp, jyamagis@nii.ac.jp}

\begin{document}

\maketitle

\begin{abstract}
Speech enhancement (SE) methods mainly focus on recovering clean speech from noisy input. In real-world speech communication, however, noises often exist in not only speaker but also listener environments. Although SE methods can suppress the noise contained in the speaker's voice, they cannot deal with the noise that is physically present in the listener side. To address such a complicated but common scenario, we investigate a deep learning-based joint framework integrating noise reduction (NR) with listening enhancement (LE), in which the NR module first suppresses noise and the LE module then modifies the denoised speech, i.e., the output of the NR module, to further improve speech intelligibility. The enhanced speech can thus be less noisy and more intelligible for listeners. Experimental results show that our proposed method achieves promising results and significantly outperforms the disjoint processing methods in terms of various speech evaluation metrics.

\end{abstract}

\section{Introduction}

\begin{figure}
    \begin{center}
        \includegraphics[width=0.471\textwidth]{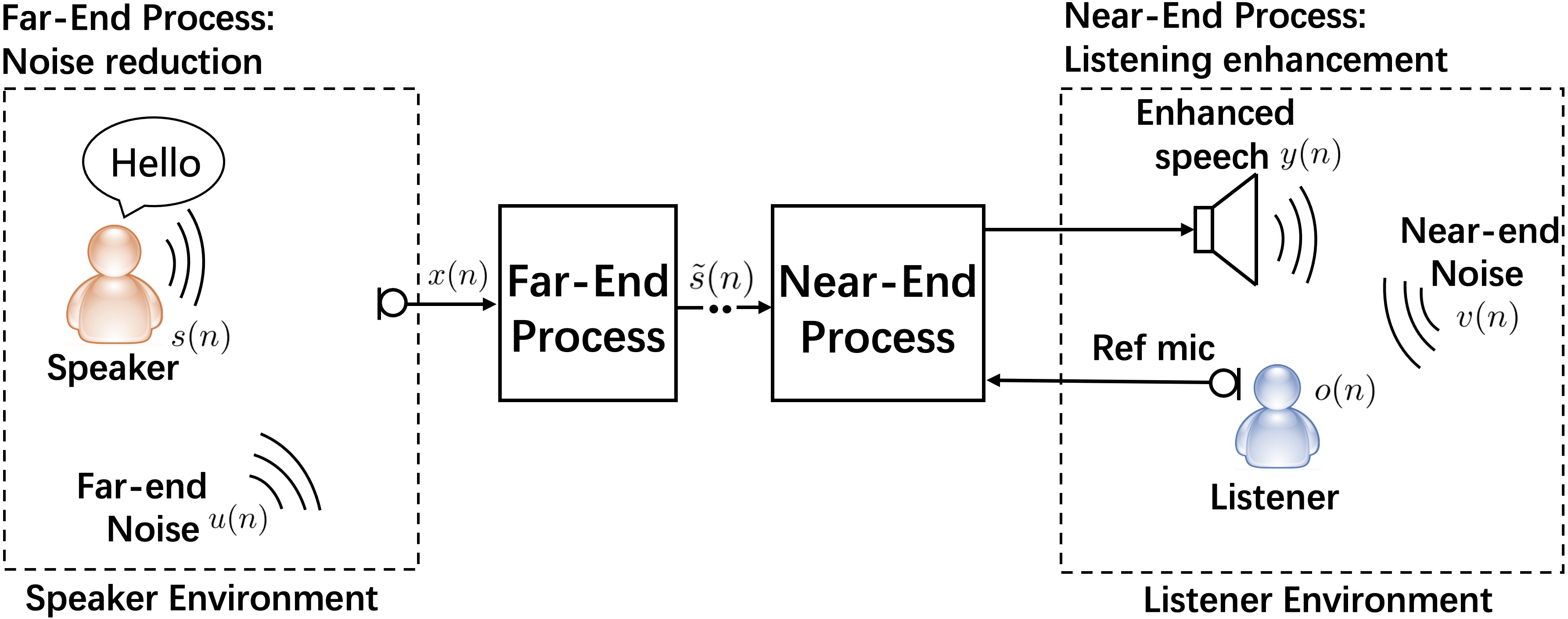}\\
        \caption{A scenario of real-world speech communication where noises exist in both speaker and listener environments. A reference microphone is used to measure the near-end noise properties.}
        \label{fig:scenario}
    \end{center}
    \vspace{-7mm}
\end{figure}

Speech communication systems, such as mobile telephony and hearing aids, are supposed to work under adverse conditions of environmental noise. In real-world application scenarios, as depicted in Fig.~\ref{fig:scenario}, noises may exist in not only far-end speaker but also near-end listener environments, resulting in severe degradation of speech quality and intelligibility. To improve the listener's listening experience, speech processing should be accordingly carried out as two sub-tasks: (1) noise reduction (NR): to suppress noise and recover clean speech in the far-end speaker side; and (2) listening enhancement (LE): to pre-process speech signals (the output of the NR module) before playback to improve its intelligibility in the near-end listener side. In this paper, we refer to this two-stage task as \textit{full-end speech enhancement}.

Many previous works have been proposed to address the above-mentioned sub-tasks separately. For the far-end NR scenario, deep neural network (DNN)-based speech enhancement (SE) methods \cite{lu2013speech, xu2014regression, weninger2015speech, fu2019metricgan} have gradually become mainstream and shown notable improvement over traditional methods \cite{boll1979suppression, ephraim1985speech}. For the near-end LE scenario, the common idea is to redistribute speech energy in the time-frequency domain in such a way as to boost the perceptually important acoustic cues. A redistribution strategy can be artificially designed on the basis of expert knowledge (e.g., spectral tilt flattening and dynamic range compression \cite{ZorilaKS12, Chermaz2020}) or automatically derived from an optimization solution for an intelligibility metric (e.g., speech intelligibility index (SII) \cite{american1997american}). In our previous work, we introduced generative adversarial network (GAN) model into LE task and achieved impressive results \cite{li2021multi}. However, these methods assume that the input speech is clean, which is incompatible with the scenario of a full-end SE task where the input speech is often noisy.

Recently, researchers have explored joint processing of noise reduction and listening enhancement \cite{niermann2017joint, khademi2017intelligibility, fuglsig2021joint, pv2021end}. 
To achieve it, most existing works jointly control the NR filter along with near-end filter gain to optimize a certain target intelligibility metric, e.g., SII in \cite{niermann2017joint} and mutual information in \cite{khademi2017intelligibility}. However, to make the optimization problem mathematically tractable, the NR filter is considered to be relatively simple (e.g., Niermann \MakeLowercase{\textit{et al.}} \cite{niermann2017joint} used Wiener filter). Besides, most of them introduce additional assumptions and approximations to some extent, including target metric approximation \cite{niermann2017joint, fuglsig2021joint} and Gaussian signal model \cite{khademi2017intelligibility}, therefore limiting performance.

In this paper, we propose a novel joint model for full-end SE. We intuitively extend our previous GAN-based LE method \cite{li2021multi} by integrating it with a mainstream DNN-based NR method, leading to a fully DNN-based solution. This model can fully benefit from the powerful modeling capabilities of neural networks. Moreover, it can be jointly optimized using a unified loss function and without being dependent on unnecessary assumptions and approximations. Our experiments indicate that the proposed model significantly improves speech quality and intelligibility and clearly outperforms the disjoint pipeline methods.

\section{Problem Formulation}
\label{sec:sec2}

Consider the application scenario of full-end SE that is depicted in Fig.~\ref{fig:scenario}. 
The signal model follows\footnote{We disregard all related room transfer functions for simplicity.}:
\begin{equation}
    x = s+u, \quad \Tilde{s} = NR(x), \quad y = LE(\Tilde{s} | v ), \quad o = y + v,
    \label{eq:signal_model}
\end{equation}
where $s$ is the clean speech, $u$ is the far-end environmental noise\footnote{We only take into account the additive noise, but it can be generalized to other degradation such as reverberation and audio clipping.}, $v$ is the near-end environmental noise, and $x$ is the signal received by the far-end microphone. The NR module receives $x$ and outputs the estimated clean speech $\Tilde{s}$, i.e., the denoised speech. By conditioning on the near-end noise estimation, the LE module further modifies $\Tilde{s}$ before it is played by loudspeaker. The output enhanced speech is denoted as $y$. 
Finally, the signal $o$ is observed by the near-end listener. Our goal is to improve the listening experience for listeners, i.e., the quality of $y$ (without the near-end noise $v$) and intelligibility of $o$ under $v$, by designing effective NR and LE modules. 
Also, to limit loudspeaker overload and unpleasant playback volume, we follow the equal-power constraint that requires that signal power before and after LE modification (i.e., $\Tilde{s}$ and $y$) to be the same. 

\begin{figure}
    \begin{center}
        \includegraphics[width=\linewidth]{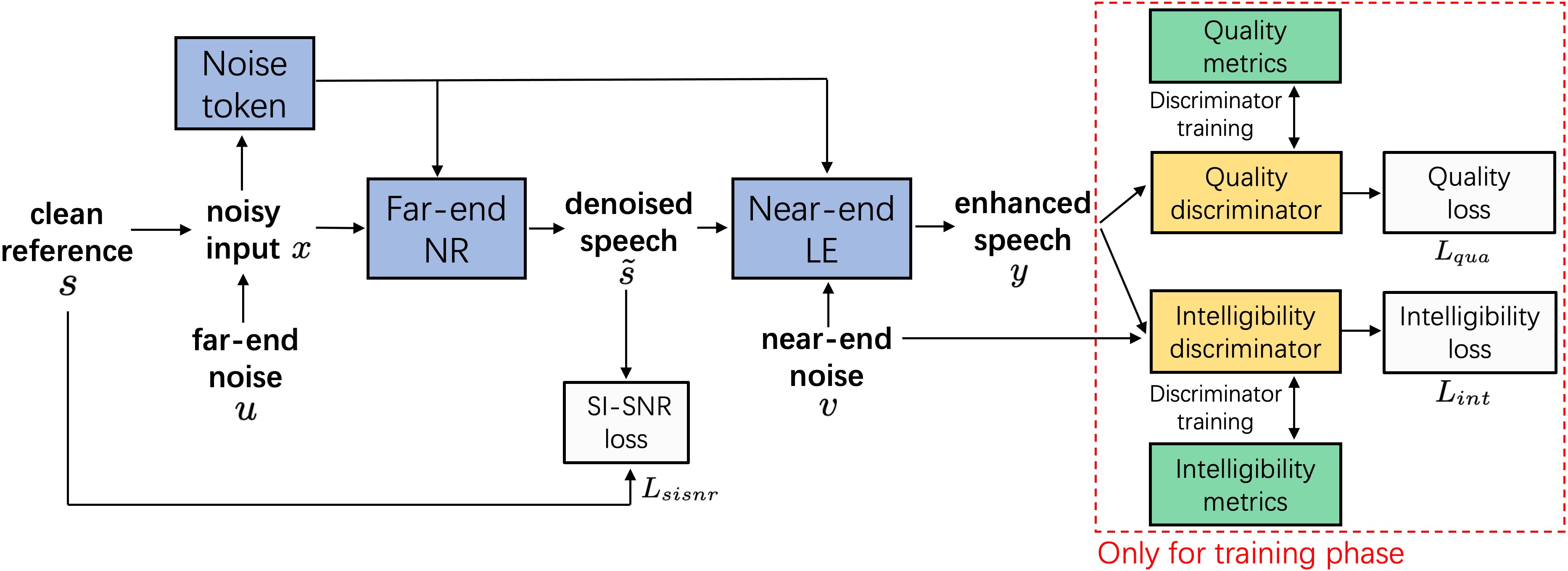}\\
        \caption{Overall diagram of the proposed method.}
        \label{fig:diagram}
    \end{center}
    \vspace{-8mm}
\end{figure}

\begin{figure*}[t]
    \centering
    \subfigure[Noise reduction module]{
        \includegraphics[height=26mm,width=53mm]{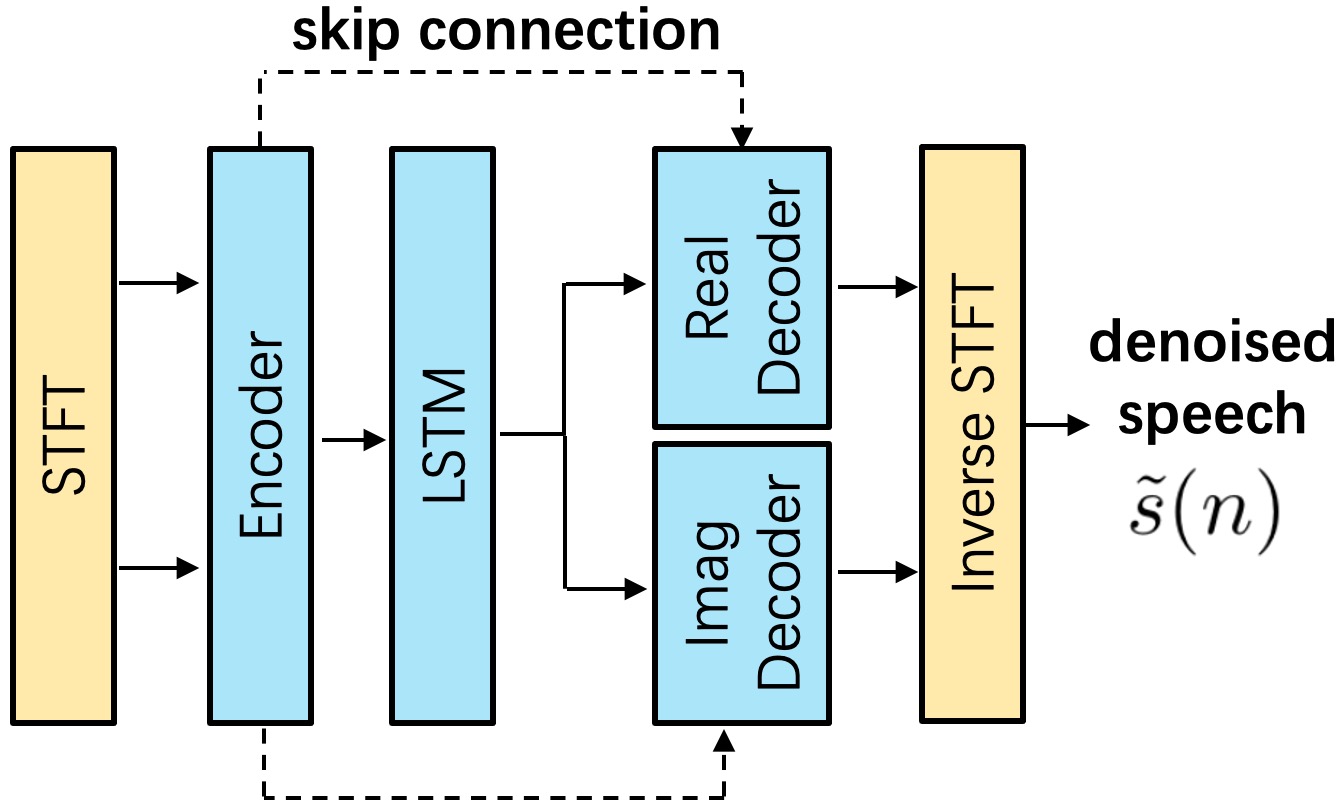}
        }
    \subfigure[Listening enhancement module]{
        \includegraphics[height=24mm,width=53mm]{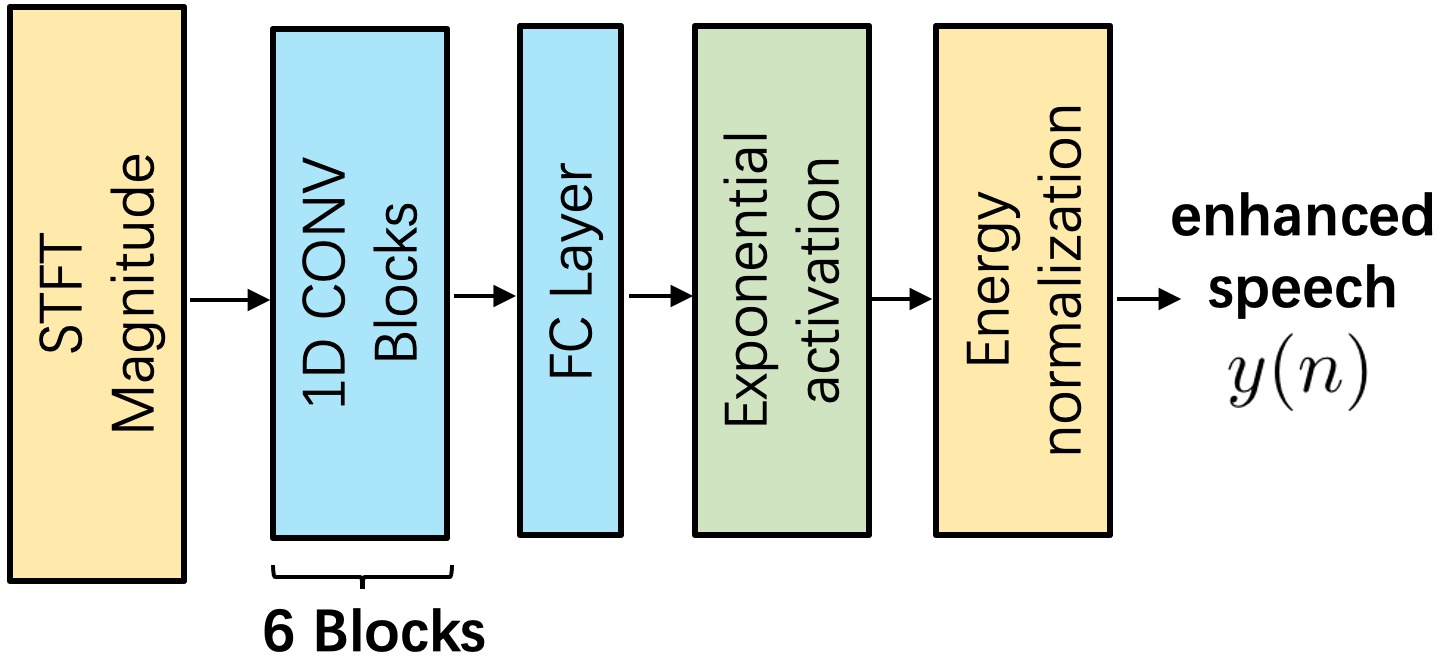}
        }     
    \subfigure[Noise token module]{
        \includegraphics[height=24mm,width=53mm]{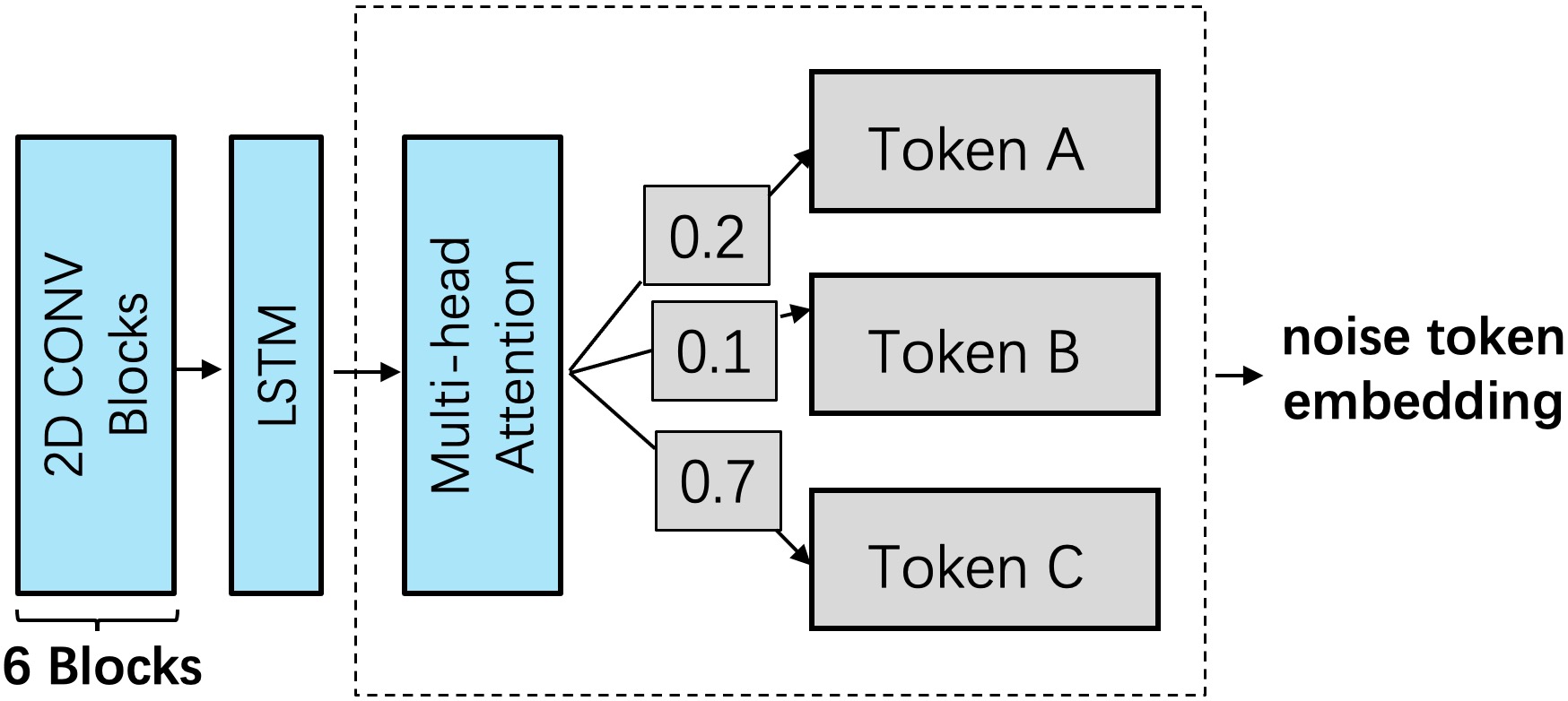}
        }    
    \caption{Illustration of the three major modules. CONV and FC denote convolutional and fully-connected layers, respectively.}
    \label{fig:details_of_modules}
    \vspace{-5.5mm}
\end{figure*}

\vspace{-1mm}
\section{Proposed Method}
\label{sec:proposed_method}
In this section, we introduce the proposed joint model. Figure~\ref{fig:diagram} shows its overall diagram. 
It consists of three main modules: (1) a far-end NR module that suppresses noise; (2) a near-end LE module that increases the intelligibility of denoised speech by redistributing its energy over time and frequency; and (3) the noise token module that extracts noise embedding and informs other modules of far-end environmental information. 
We use causal configurations for these three modules, which enables the model to perform real-time speech processing.
Next, we will describe details of each module.

\vspace{-1mm}
\subsection{Far-end Noise Reduction}
Far-end NR aims to suppress noise. The input is the noisy speech recorded by the far-end microphone, and the ideal output is a clean speech signal without noise disturbance.

To achieve this, we use a convolutional recurrent network (CRN) \cite{tan2019complex} as the main neural architecture. As shown in Fig.~\ref{fig:details_of_modules}(a), the noisy speech is first converted into real and imaginary spectrograms by using short-time Fourier transformation (STFT). We use a Hanning window with window size of 32 ms and hop size of 8 ms. The encoder consists of five 2D convolutional layers each with 1 (along the time axis)$\times$3 (along the frequency axis) kernel, 1$\times$2 stride, layer normalization \cite{ba2016layer}, and parametric ReLU (PReLU). The output channels are set to 16, 32, 48, 64, 96, and 128, respectively. Between the encoder and the decoder, we insert a two-layer unidirectional long short-term memory (LSTM) with 512 nodes to model the temporal dependencies. The decoder comprises five transposed 2D convolutional layers with the same kernel and stride size as the encoder. Since skip connections are used to feed the output of each encoder layer as the additional input of the decoder layer, the output channels of the decoder are accordingly set to 256, 192, 128, 96, 64, and 32, respectively. The following two decoders respectively predict the real and imaginary parts of a complex ratio mask \cite{williamson2015complex}, which are then multiplied with the original complex spectrogram to obtain the denoised one. The denoised speech is then generated with inverse STFT and passed on to the following LE module.

\subsection{Near-end Listening Enhancement}
\label{sec:nele}
The LE module modifies the denoised speech to make it sound more intelligible under the near-end environmental noise.
Unlike the far-end process in which the clean speech is ground-truth labels, in the near-end case, there is no definition of what the \textit{perfectly intelligible} speech is. 
To address this, in our previous work \cite{li2021multi}, 
we modified speech in such a way as to maximize the selected speech intelligibility metrics.
Moreover, we also incorporated multiple speech quality metrics as optimization targets to compensate for the quality loss caused by intelligibility-enhancing modifications. The target metrics are SIIB \cite{van2017instrumental}, HASPI \cite{kates2021hearing}, ESTOI \cite{jensen2016algorithm}, PESQ \cite{rix2001perceptual}, ViSQOL \cite{chinen2020visqol}, and HASQI \cite{kates2021hearing}. The former three are intelligibility metrics and the latter three are quality metrics. 

However, these metrics are quite complex and even non-differentiable, resulting in difficulty in optimization. Therefore, we introduced GAN model into the LE task, in which the generator ($G$) enhances the intelligibility of speech and the discriminator ($D$) mimics the behavior of objective metrics by learning to predict the correct intelligibility scores of modified speech. Specifically, the discriminator has to closely approximate the intelligibility metrics as a learned surrogate, and then the generator can be trained with the guidance of this surrogate. 
As shown in Fig.~\ref{fig:diagram}, the input features for the LE module ($G$) includes: (1) spectrogram magnitude of denoised speech, (2) the near-end noise estimation (i.e., noise power spectral density estimated by reference microphone), and (3) neural noise embedding extracted from noisy input speech (we will explain this in Section~\ref{sec: noise_token}).
The architecture of $G$ is shown in Fig.~\ref{fig:details_of_modules}(b). It consists of six 1D convolutional layers each with cumulative layer normalization \cite{luo2019conv} and PReLU activation, followed by an FC layer. The detailed parameters are same as those used in \cite{li2021multi}.
The element-wise exponential activation is given as follow:
\begin{equation}
    \label{eq:exponential}
    \alpha = exp(4*tanh(u)),
\end{equation}
where $u$ is the result of the last FC layer. It predicts the amplification factors $\alpha$ ranging from 0.02 to 55. The $\alpha$ is then applied to the spectrogram of denoised speech to redistribute its energy across time and frequency bins: the bin is boosted when $\alpha > 1$ and otherwise suppressed.
Finally, we apply an energy normalization layer to satisfy the equal-power constraint. 

For discriminator ($D$), we prepare $D_{qua}$ and $D_{int}$ for predicting the quality and intelligibility scores of enhanced speech, respectively. 
Each discriminator consists of five 2D convolutional layers with the same parameters used in \cite{li2021multi}.
The sigmoid activation acts as the final output layer of discriminators, which predicts the scores of modelled target metrics.
For example, the output nodes of $D_{int}$ are set to 3, corresponding to three target intelligibility metrics, i.e., SIIB, HASPI, and ESTOI.

\vspace{-1mm}
\subsection{Noise Token}
\label{sec: noise_token}
\vspace{-0.5mm}
We also insert the noise token module \cite{li2020noise} into the joint model. Noise tokens are a set of neural noise templates used to encode the far-end environment information and generate the noise embedding. Such embedding is regarded as additional noise knowledge and fed into both NR and LE modules. We previously demonstrated \cite{li2020noise} that noise token embedding can improve the performance of the NR module. We expect that they can also benefit the LE module. For example, by exploiting far-end noise knowledge, the LE module may avoid amplifying those time-frequency regions containing much residual noise.

Fig.~\ref{fig:details_of_modules}(c) shows the detailed structure of the noise token module. The input feature is noisy spectrogram magnitude. It is composed of six 2D convolutional layers each with 3$\times$3 kernel and 1$\times$2 stride. The channels are set to 32, 32, 64, 64, 128, and 128, respectively. A uni-directional LSTM layer with 256 nodes is followed by the last CONV block, resulting in a 256-dimensional encoded representation. Next, in the multi-head attention model \cite{vaswani2017attention}, this representation serves as the \textit{query} and 16 trainable 256-dimensional noise tokens (representing different noise patterns) serve as the \textit{key} and \textit{vector}. We set the number of attention heads to 8. Finally, the noise token embedding can be generated as the weighted sum of the noise tokens.

\vspace{-1.5mm}
\subsection{Training Objective}
\label{sec:training_obj}
The training objective is composed of three terms:
\begin{equation}
    \label{eq:training_obj}
    L = L_{int} + \alpha*L_{qua} + \beta*L_{sisnr},
\end{equation}
where $L_{int}$ is intelligibility loss calculated by the intelligibility discriminator, $L_{qua}$ is quality loss calculated by the quality discriminator, and $L_{sisnr}$ is speech denoising loss. $\alpha$ and $\beta$ denote the weight parameters, respectively. To be more specific, $L_{sisnr}$ is the scale-invariant signal-to-noise ratio (SI-SNR) \cite{luo2019conv}, which is calculated by comparing the denoised speech with the clean reference speech.
Intelligibility loss $L_{int}$ is defined as the mean square error between the predicted intelligibility scores and the maximum scores of target metrics:
\begin{equation}
    L_{int} = ||D_{int}(y|v) - t_{int}||^2\\
    \label{eq:loss_L}
\end{equation}
where $y$ is the final enhanced speech output by the LE module, $D_{int}(y|v)$ is the predicted scores (under noise $v$) output by the intelligibility discriminator, and $t_{int}$ is the maximum scores of the selected intelligibility metrics, respectively. By means of this loss, the LE module has to reach intelligibility scores as high as possible. Similarly we can define the quality loss $L_{qua}$. We jointly optimize the whole model (including noise token, NR, and LE modules) by using the loss function of Eq.~(\ref{eq:training_obj}).

\begin{table*}[t]
    \caption{Objective intelligibility scores averaged over three near-end SNRs for each far-end SNR condition.}
    \vspace{-1.5mm}
    \label{tab:int}
    \centering
    \renewcommand\tabcolsep{3.6pt}
    \renewcommand\arraystretch{1.2}
    \scalebox{0.896}{
    \begin{tabular}{p{64pt}<{\centering} p{28.5pt}<{\centering}p{28.5pt}<{\centering}p{28.5pt}<{\centering} c p{28.5pt}<{\centering}p{28.5pt}<{\centering}p{28.5pt}<{\centering} c p{28.5pt}<{\centering}p{28.5pt}<{\centering}p{28.5pt}<{\centering}}
        \hline
        \hline
           \multirow{2}{*}{System} &
           \multicolumn{3}{c}{Far-end SNR $=6$ dB} & & \multicolumn{3}{c}{Far-end SNR $=10$ dB} & &
           \multicolumn{3}{c}{Far-end SNR $=14$ dB} \\
           \cline{2-4}  \cline{6-8} \cline{10-12} 
            & SIIB & HASPI & ESTOI & & SIIB & HASPI & ESTOI & & SIIB & HASPI & ESTOI \\
           \hline
          Noisy & 17.98 & 2.20 & 0.221 & & 19.72 & 2.31 & 0.237 & & 21.07 & 2.41 & 0.249  \\
          Noisy+NR & 19.52 & 2.24 & 0.250 & & 20.73 & 2.32 & 0.259 & & 21.65 & 2.39 & 0.266  \\
          Noisy+LE & 15.79 & 2.09 & 0.180 & & 18.76 & 2.28 & 0.206 & & 21.91 & 2.47 & 0.232  \\
        \hline
          DSPPipe & 15.58 & 1.96 & 0.208 &  & 18.22 & 2.10 & 0.229 & & 21.06 & 2.24 & 0.251 \\
          NeuralPipe & 24.47 & 2.67 & 0.302 & & 27.34 & 2.85 & 0.319 & & 30.09 & \textbf{3.00} & 0.333  \\
          Joint & 26.16 & 2.70 & 0.305 & & 28.65 & 2.84 & 0.319 & & 30.77 & 2.96 & 0.330  \\
          Joint+NT & \textbf{28.48} & \textbf{2.73} & \textbf{0.320} & & \textbf{31.45} & \textbf{2.87} & \textbf{0.334} & & \textbf{33.79} & 2.99 & \textbf{0.344}  \\
        \hline
        \hline
    \end{tabular}
    }
    \vspace{-2mm}
\end{table*}

\begin{table*}[t]
    \caption{Objective quality scores averaged over three near-end SNRs for each far-end SNR condition.}
    \vspace{-1.5mm}
    \label{tab:qua}
    \centering
    \renewcommand\tabcolsep{3pt}
    \renewcommand\arraystretch{1.2}
    \scalebox{0.896}{
    \begin{tabular}{p{64pt}<{\centering} p{28.5pt}<{\centering}p{28.5pt}<{\centering}p{32pt}<{\centering} c p{28.5pt}<{\centering}p{28.5pt}<{\centering}p{32pt}<{\centering} c p{28.5pt}<{\centering}p{28.5pt}<{\centering}p{32pt}<{\centering}}
        \hline
        \hline
           \multirow{2}{*}{System} &
           \multicolumn{3}{c}{Far-end SNR $=6$ dB} & & \multicolumn{3}{c}{Far-end SNR $=10$ dB} & &
           \multicolumn{3}{c}{Far-end SNR $=14$ dB} \\
           \cline{2-4}  \cline{6-8} \cline{10-12} 
            & PESQ & HASQI & ViSQOL & & PESQ & HASQI & ViSQOL & & PESQ & HASQI & ViSQOL \\
           \hline
          Noisy & 1.41 & 0.15 & 1.83 & & 1.55 & 0.18 & 1.94 & & 1.69 & 0.21 & 2.09  \\
          Noisy+NR & \textbf{2.33} & 0.28 & \textbf{2.48} & & \textbf{2.52} & 0.32 & \textbf{2.69} & & \textbf{2.70} & \textbf{0.36} & \textbf{2.91}  \\
          Noisy+LE & 1.24 & 0.10 & 1.66 & & 1.32 & 0.12 & 1.71 & & 1.41 & 0.14 & 1.78  \\
        \hline
          DSPPipe & 1.32 & 0.10 & 1.68 & & 1.43 & 0.12 & 1.74 & & 1.54 & 0.14 & 1.81 \\
          NeuralPipe & 2.01 & 0.23 & 2.14 & & 2.19 & 0.26 & 2.25 & & 2.35 & 0.28 & 2.35  \\
          Joint & 2.14 & 0.28 & 2.20 & & 2.30 & 0.30 & 2.32 & & 2.43 & 0.33 & 2.43  \\
          Joint+NT & 2.26 & \textbf{0.30} & 2.32 & & 2.45 & \textbf{0.32} & 2.43 & & 2.58 & 0.35 & 2.52  \\
        \hline
        \hline
    \end{tabular}
    }
    \vspace{-3mm}
\end{table*}

\vspace{-1.5mm}
\section{Experiments}
\label{sec:experiment}
\vspace{-0.5mm}
\subsection{Data Preparation}
\vspace{-0.5mm}
\label{sec:data_prepare}
We used two public corpora of Harvard sentences \cite{rothauser1969ieee} (one spoken by male \cite{valentini2019hurricane} and one by female \cite{demonte2019}) in the experiments. We split the whole 720 Harvard sentences into 600, 60, and 60 for training, validation, and test data, respectively.

For training and validation, eight noise types were used in both far-end (speaker) and near-end (listener) environments.
Far-end SNR levels were set to 4, 8, and 12 dB; near-end SNR levels were set to -11, -7, and -3 dB. By randomly combining these settings, we generated 28,800 and 2,880 utterances for training and validation, respectively.

For test set, the far-end noise type is cafeteria at three SNRs, i.e., 6, 10, and 14 dB; near-end noise type is airport announcement at three SNRs, i.e., -9, -5, and -1 dB. To summarize, the test set contained 1,080 utterances (60 sentences $\times$ 2 genders $\times$ 3 far-end SNRs $\times$ 3 near-end SNRs). Note that all the sentences, SNR levels, and noise types of the test set were unseen during model training.

\vspace{-1.5mm}
\subsection{Implementation Details}
\label{sec:implementation_details}
\vspace{-0.5mm}

All signals used in the experiments were resampled at 16 kHz. Improved minima controlled recursive averaging algorithm (IMCRA) \cite{cohen2003noise} was used to estimate power spectral density of the near-end noise.
During training, we applied parametric logistic function to normalize all metric scores into the range of $[0, 1]$, i.e., the same range with sigmoid activation, and set the corresponding target maximum scores (e.g., $t_{int}$ in Eq.~(\ref{eq:loss_L})) to 1.
We used Adam optimizer \cite{kingma2014adam} for training, with initial learning rates of
0.0002 for the three neural module components (noise token, NR, and LE) and 0.0001 for the discriminators ($D_{int}$ and $D_{qua}$).
The batch size was 1, and the hyper-parameters $\alpha$ and $\beta$ in Eq.~(\ref{eq:training_obj}) were set to 0.6 and 0.005, respectively.

\vspace{-1mm}
\subsection{Objective Evaluations}
\label{sec:obj_results}
\vspace{-0.5mm}
We evaluated the proposed method using six objective metrics. As mentioned in Section~\ref{sec:nele}, the intelligibility metrics are SIIB, HASPI, and ESTOI; the quality metrics are PESQ, ViSQOL, and HASQI. For the above-mentioned metrics, higher scores indicate better performance. The far-end noisy speech is processed by a certain system and then played under the near-end noise. We evaluated seven systems\footnote{Audio samples: \url{https://nii-yamagishilab.github.io/hyli666-demos/full-end-se}} and notate them as follows.
\begin{itemize}
    \vspace{-1mm}
    \item \textbf{Noisy}: The far-end input noisy speech is played under the near-end noise without any modification.
    \vspace{-1mm}
    \item \textbf{Noisy+NR}: The far-end input noisy speech is processed only by the NR module.
    \vspace{-1mm}
    \item \textbf{Noisy+LE}: Processed only by the LE module.
    \vspace{-1mm}
    \item \textbf{DSPPipe}: Processed by the signal processing-based disjoint pipeline, which consists of Wiener filter (for NR) and SSDRC algorithm \cite{ZorilaKS12} (for LE). 
    \vspace{-1mm}
    \item \textbf{NeuralPipe}: Processed by neural network-based disjoint pipeline, which consists of the pretrained CRN-based NR \cite{tan2019complex} and GAN-based LE \cite{li2021multi} modules. 
    \vspace{-1mm}
    \item \textbf{Joint}: Processed by the partial joint model (without the noise token module), in which the NR and LE models are jointly optimized.
    \vspace{-1mm}
    \item \textbf{Joint+NT}: Processed by the full proposed joint model (with the noise token module).
\end{itemize}

Intelligibility evaluation results are listed in Table~\ref{tab:int}, where the scores were averaged over the three near-end SNR levels. As we can see, applying only NR (\textbf{Noisy+NR}) or LE (\textbf{Noisy+LE}) does not increase the intelligibility. \textbf{Noisy+LE} has even lower scores than \textbf{Noisy}, since the LE module amplifies the noise contained in the noisy input. To address the full-end SE problem, \textbf{NeuralPipe}, \textbf{Joint}, and \textbf{Joint+NT} integrate both NR and LE modules, resulting in significant intelligibility gains compared with \textbf{Noisy}. In contrast, \textbf{DSPPipe} has extremely low scores. This is because the SSDRC processor wrongly amplifies the residual noise that is produced by the former Wiener filter.
Besides, we can clearly see that joint trained models improve upon the disjoint processing methods (\textbf{DSPPipe} and \textbf{NeuralPipe}). Moreover, benefiting from the noise token module that exploits the far-end environment information, \textbf{Joint+NT} consistently outperforms \textbf{Joint} and achieves the overall best performance.

Table~\ref{tab:qua} lists the objective quality scores of enhanced speech $y$ (without the near-end noise $v$). Since intelligibility-enhancing modifications inevitably degrade the speech quality at the cost of increasing intelligibility, \textbf{Noisy+NR} performs better than the proposed joint models. However, joint models preserve speech quality much better than \textbf{DSPPipe} and \textbf{NeuralPipe}, which indicates the effectiveness of our proposed method. 

\begin{figure}[t]
    \centering
    \subfigure[Intelligibility preference test]{
        \includegraphics[height=24.3mm,width=73.5mm]{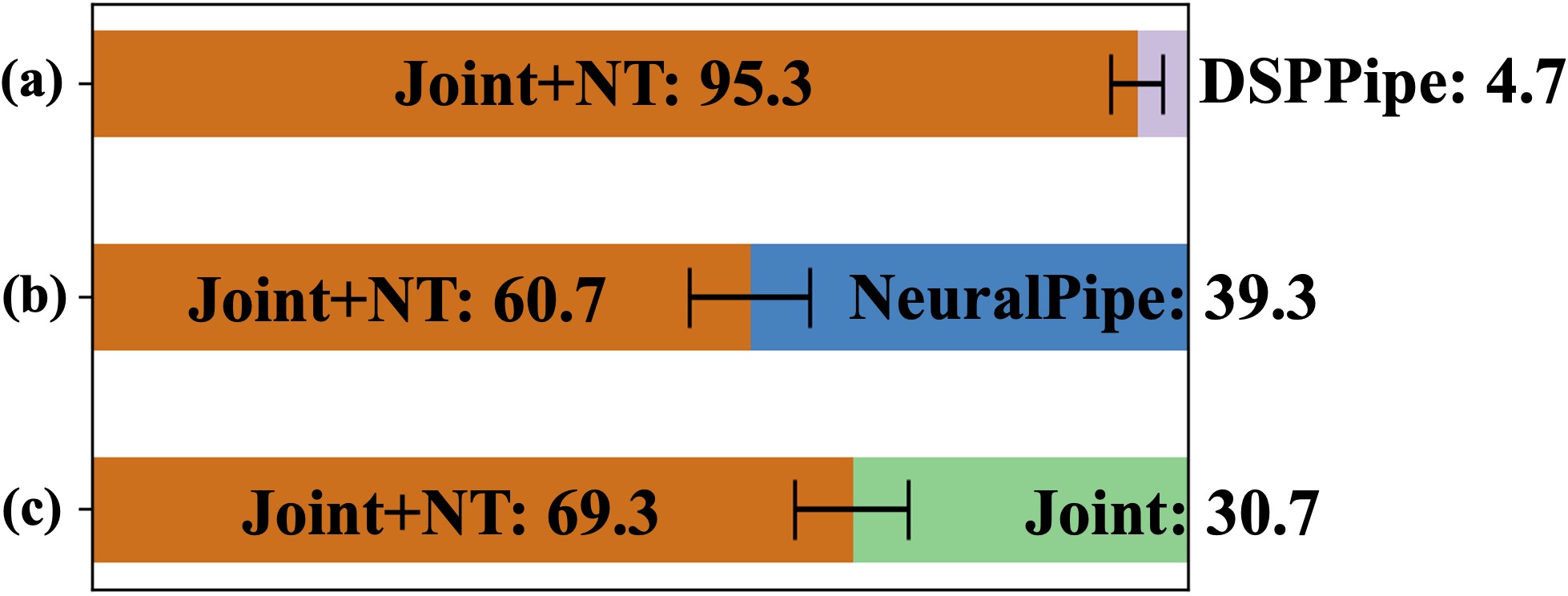}
        }
        \vspace{-1.5mm}
    \subfigure[Quality preference test]{
        \includegraphics[height=24.3mm,width=73.5mm]{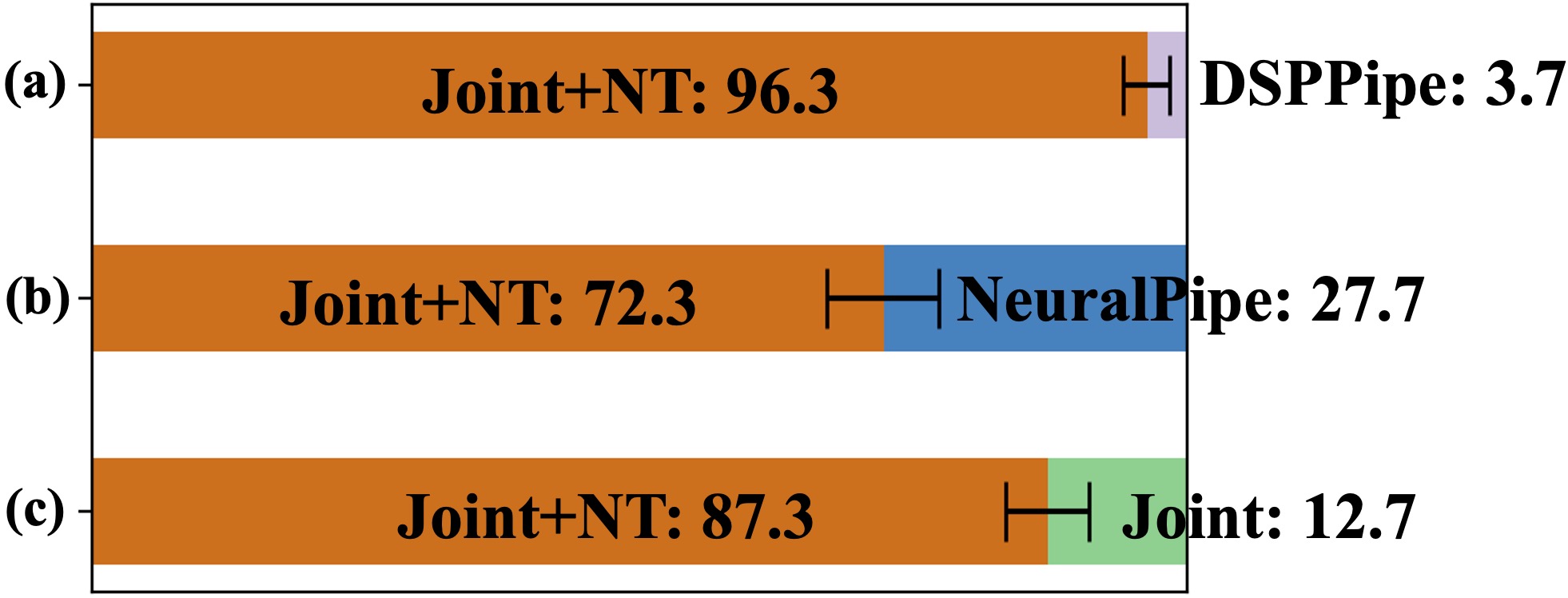}
        }     
        \vspace{-1.5mm}
    \caption{Preference scores (\%) with 95\% confidence intervals.}
    \vspace{-5mm}
    \label{fig:subjective_result}
\end{figure}

\vspace{-1.5mm}
\subsection{Subjective Listening Tests}
\label{sec:sub_results}
\vspace{-0.5mm}
We conducted subjective preference tests to further evaluate the speech intelligibility and perceptual quality.
We conducted pairwise comparisons between \mbox{\textbf{Joint+NT}} and the three systems: (1) \mbox{\textbf{DSPPipe}}, (2) \mbox{\textbf{NeuralPipe}}, and (3) \mbox{\textbf{Joint}}.
300 enhanced samples were randomly selected from the test set for each system, resulting in 900 tested sample pairs (300 samples $\times$ 3 system pairs). A total of 20 native English speakers were recruited to participate in intelligibility and quality preference tests, respectively, and all were paid. 
For intelligibility test, each participant was instructed to listen to 45 randomized sample pairs played back under the near-end noise (i.e., the signal $o$ in Eq.~(\ref{eq:signal_model})), and for each pair, they had to select the one that sounded clearer or that they could hear with less listening efforts. For quality test, each participant had to listen to 45 sample pairs (without the near-end noise, i.e., the signal $y$ in Eq.~(\ref{eq:signal_model})) and select the one that sounded better in terms of listening quality.
As we can see from Fig.~\ref{fig:subjective_result}, the proposed \textbf{Joint+NT} achieved significantly higher preference scores than all three compared systems in terms of both speech intelligibility and quality. More interestingly, we found that \textbf{Joint+NT} outperformed \textbf{Joint} by a large margin in listening test results, which further indicates that exploiting far-end noise knowledge is crucial for not only far-end noise reduction but also near-end listening enhancement.

\vspace{-1.5mm}
\section{Conclusion}
\label{sec:conclusion}
\vspace{-0.5mm}
To address the full-end speech enhancement task where both speaker and listener environments are noisy, we proposed a DNN-based joint framework integrating noise reduction with listening enhancement. These two modules can be jointly optimized under the proposed framework. The NR module suppresses the noise of the input noisy speech, and the LE module further improves its intelligibility. 
Experimental results using both objective evaluations and subjective listening tests indicate that the joint model can achieve significant intelligibility gain while preserving speech quality well. It also consistently outperforms the disjoint processing pipelines by a large margin. 

\vspace{0.5mm}

\noindent
\textbf{Acknowledgments}
This work was partially supported by a JST CREST Grant (JPMJCR18A6, VoicePersonae project), Japan, partially by MEXT KAKENHI Grants, 18H04112 and 21H04906, Japan, and partially by SOKENDAI, Japan.

\bibliography{template.bbl}

\begin{thebibliography}{10}
\providecommand{\url}[1]{#1}
\csname url@samestyle\endcsname
\providecommand{\newblock}{\relax}
\providecommand{\bibinfo}[2]{#2}
\providecommand{\BIBentrySTDinterwordspacing}{\spaceskip=0pt\relax}
\providecommand{\BIBentryALTinterwordstretchfactor}{4}
\providecommand{\BIBentryALTinterwordspacing}{\spaceskip=\fontdimen2\font plus
\BIBentryALTinterwordstretchfactor\fontdimen3\font minus
  \fontdimen4\font\relax}
\providecommand{\BIBforeignlanguage}[2]{{%
\expandafter\ifx\csname l@#1\endcsname\relax
\typeout{** WARNING: IEEEtran.bst: No hyphenation pattern has been}%
\typeout{** loaded for the language `#1'. Using the pattern for}%
\typeout{** the default language instead.}%
\else
\language=\csname l@#1\endcsname
\fi
#2}}
\providecommand{\BIBdecl}{\relax}
\BIBdecl

\bibitem{lu2013speech}
X.~Lu, Y.~Tsao, S.~Matsuda, and C.~Hori, ``Speech enhancement based on deep
  denoising autoencoder.'' in \emph{Interspeech}, 2013, pp. 436--440.

\bibitem{xu2014regression}
Y.~Xu, J.~Du, L.-R. Dai, and C.-H. Lee, ``A regression approach to speech
  enhancement based on deep neural networks,'' \emph{IEEE/ACM Transactions on
  Audio, Speech, and Language Processing}, vol.~23, no.~1, pp. 7--19, 2014.

\bibitem{weninger2015speech}
F.~Weninger, H.~Erdogan, S.~Watanabe, E.~Vincent, J.~Le~Roux, J.~R. Hershey,
  and B.~Schuller, ``{Speech enhancement with {LSTM} recurrent neural networks
  and its application to noise-robust ASR},'' in \emph{International Conference
  on Latent Variable Analysis and Signal Separation}.\hskip 1em plus 0.5em
  minus 0.4em\relax Springer, 2015, pp. 91--99.

\bibitem{fu2019metricgan}
S.-W. Fu, C.-F. Liao, Y.~Tsao, and S.-D. Lin, ``{MetricGAN: Generative
  adversarial networks based black-box metric scores optimization for speech
  enhancement},'' in \emph{International Conference on Machine Learning}.\hskip
  1em plus 0.5em minus 0.4em\relax PMLR, 2019, pp. 2031--2041.

\bibitem{boll1979suppression}
S.~Boll, ``Suppression of acoustic noise in speech using spectral
  subtraction,'' \emph{IEEE Transactions on Acoustics, Speech, and Signal
  Processing}, vol.~27, no.~2, pp. 113--120, 1979.

\bibitem{ephraim1985speech}
Y.~Ephraim and D.~Malah, ``Speech enhancement using a minimum mean-square error
  log-spectral amplitude estimator,'' \emph{IEEE Transactions on Acoustics,
  Speech, and Signal Processing}, vol.~33, no.~2, pp. 443--445, 1985.

\bibitem{ZorilaKS12}
T.-C. Zorila, V.~Kandia, and Y.~Stylianou, ``Speech-in-noise intelligibility
  improvement based on spectral shaping and dynamic range compression,'' in
  \emph{Proc. Interspeech}, 2012, pp. 635--638.

\bibitem{Chermaz2020}
C.~Chermaz and S.~King, ``{A Sound Engineering Approach to Near End Listening
  Enhancement},'' in \emph{Proc. Interspeech}, 2020, pp. 1356--1360.

\bibitem{american1997american}
A.~N.~S. Institute, \emph{{American National Standard: Methods for Calculation
  of the Speech Intelligibility Index}}.\hskip 1em plus 0.5em minus 0.4em\relax
  Acoustical Society of America, 1997.

\bibitem{li2021multi}
H.~Li and J.~Yamagishi, ``Multi-metric optimization using generative
  adversarial networks for near-end speech intelligibility enhancement,''
  \emph{IEEE/ACM Transactions on Audio, Speech, and Language Processing},
  vol.~29, pp. 3000--3011, 2021.

\bibitem{niermann2017joint}
M.~Niermann, P.~Jax, and P.~Vary, ``Joint near-end listening enhancement and
  far-end noise reduction,'' in \emph{2017 IEEE International Conference on
  Acoustics, Speech and Signal Processing (ICASSP)}.\hskip 1em plus 0.5em minus
  0.4em\relax IEEE, 2017, pp. 4970--4974.

\bibitem{khademi2017intelligibility}
S.~Khademi, R.~C. Hendriks, and W.~B. Kleijn, ``Intelligibility enhancement
  based on mutual information,'' \emph{IEEE/ACM Transactions on Audio, Speech,
  and Language Processing}, vol.~25, no.~8, pp. 1694--1708, 2017.

\bibitem{fuglsig2021joint}
A.~J. Fuglsig, J.~{\O}stergaard, J.~Jensen, L.~S. Bertelsen, P.~Mariager, and
  Z.-H. Tan, ``Joint far-and near-end speech intelligibility enhancement based
  on the approximated speech intelligibility index,'' \emph{arXiv preprint
  arXiv:2111.07759}, 2021.

\bibitem{pv2021end}
M.~P. Shifas, T.-C. Zorila, and Y.~Stylianou, ``End-to-end neural based
  modification of noisy speech for speech-in-noise intelligibility
  improvement,'' \emph{IEEE/ACM Transactions on Audio, Speech, and Language
  Processing}, vol.~30, pp. 162--173, 2022.

\bibitem{tan2019complex}
K.~Tan and D.~Wang, ``Complex spectral mapping with a convolutional recurrent
  network for monaural speech enhancement,'' in \emph{2019 IEEE International
  Conference on Acoustics, Speech and Signal Processing (ICASSP)}.\hskip 1em
  plus 0.5em minus 0.4em\relax IEEE, 2019, pp. 6865--6869.

\bibitem{ba2016layer}
J.~L. Ba, J.~R. Kiros, and G.~E. Hinton, ``Layer normalization,'' \emph{arXiv
  preprint arXiv:1607.06450}, 2016.

\bibitem{williamson2015complex}
D.~S. Williamson, Y.~Wang, and D.~Wang, ``Complex ratio masking for monaural
  speech separation,'' \emph{IEEE/ACM transactions on audio, speech, and
  language processing}, vol.~24, no.~3, pp. 483--492, 2015.

\bibitem{van2017instrumental}
S.~Van~Kuyk, W.~B. Kleijn, and R.~C. Hendriks, ``An instrumental
  intelligibility metric based on information theory,'' \emph{IEEE Signal
  Processing Letters}, vol.~25, no.~1, pp. 115--119, 2017.

\bibitem{kates2021hearing}
J.~M. Kates and K.~H. Arehart, ``The hearing-aid speech perception index
  (haspi) version 2,'' \emph{Speech Communication}, vol. 131, pp. 35--46, 2021.

\bibitem{jensen2016algorithm}
J.~Jensen and C.~H. Taal, ``An algorithm for predicting the intelligibility of
  speech masked by modulated noise maskers,'' \emph{IEEE/ACM Transactions on
  Audio, Speech, and Language Processing}, vol.~24, no.~11, pp. 2009--2022,
  2016.

\bibitem{rix2001perceptual}
A.~W. Rix, J.~G. Beerends, M.~P. Hollier, and A.~P. Hekstra, ``{Perceptual
  evaluation of speech quality (PESQ)-a new method for speech quality
  assessment of telephone networks and codecs},'' in \emph{2001 IEEE
  International Conference on Acoustics, Speech, and Signal Processing
  (ICASSP)}, vol.~2.\hskip 1em plus 0.5em minus 0.4em\relax IEEE, 2001, pp.
  749--752.

\bibitem{chinen2020visqol}
M.~Chinen, F.~S. Lim, J.~Skoglund, N.~Gureev, F.~O'Gorman, and A.~Hines,
  ``{ViSQOL v3: An open source production ready objective speech and audio
  metric},'' in \emph{2020 Twelfth International Conference on Quality of
  Multimedia Experience (QoMEX)}.\hskip 1em plus 0.5em minus 0.4em\relax IEEE,
  2020, pp. 1--6.

\bibitem{luo2019conv}
Y.~Luo and N.~Mesgarani, ``{Conv-TasNet: Surpassing Ideal Time–Frequency
  Magnitude Masking for Speech Separation},'' \emph{IEEE/ACM Transactions on
  Audio, Speech, and Language Processing}, vol.~27, no.~8, pp. 1256--1266,
  2019.

\bibitem{li2020noise}
H.~Li and J.~Yamagishi, ``{Noise Tokens: Learning Neural Noise Templates for
  Environment-Aware Speech Enhancement},'' in \emph{Proc. Interspeech}, 2020,
  pp. 2452--2456.

\bibitem{vaswani2017attention}
A.~Vaswani, N.~Shazeer, N.~Parmar, J.~Uszkoreit, L.~Jones, A.~N. Gomez,
  {\L}.~Kaiser, and I.~Polosukhin, ``Attention is all you need,'' in
  \emph{Advances in neural information processing systems}, 2017, pp.
  5998--6008.

\bibitem{rothauser1969ieee}
E.~Rothauser, ``{IEEE recommended practice for speech quality measurements},''
  \emph{IEEE Transactions on Audio and Electroacoustics}, vol.~17, pp.
  225--246, 1969.

\bibitem{valentini2019hurricane}
\BIBentryALTinterwordspacing
C.~Valentini-Botinhao, C.~Mayo, and M.~Cooke, ``Hurricane natural speech corpus
  - higher quality version,'' 2019. [Online]. Available:
  \url{https://doi.org/10.7488/ds/2482}
\BIBentrySTDinterwordspacing

\bibitem{demonte2019}
\BIBentryALTinterwordspacing
P.~Demonte, ``{HARVARD speech corpus - audio recording 2019},'' 2019. [Online].
  Available: \url{https://doi.org/10.17866/rd.salford.c.4437578.v1}
\BIBentrySTDinterwordspacing

\bibitem{cohen2003noise}
I.~Cohen, ``{Noise spectrum estimation in adverse environments: Improved minima
  controlled recursive averaging},'' \emph{IEEE Transactions on Speech and
  Audio Processing}, vol.~11, no.~5, pp. 466--475, 2003.

\bibitem{kingma2014adam}
D.~P. Kingma and J.~Ba, ``{Adam: A method for stochastic optimization},''
  \emph{arXiv preprint arXiv:1412.6980}, 2014.

\end{thebibliography}

\end{document}